\documentclass[3p, nopreprintline, 10pt]{elsarticle}
\usepackage{lineno,hyperref}
\modulolinenumbers[5]
\usepackage{graphicx}
\usepackage{siunitx}
\usepackage{amsmath}
\usepackage{braket}
\usepackage{booktabs}
\usepackage{txfonts}
\usepackage[T1]{fontenc}
\journal{ }

\begin{document}

\begin{frontmatter}
\title{Unfolding the Hong-Ou-Mandel interference \\ between heralded photons from narrowband twin beams}
\author[ulm]{K.~Laiho\corref{mycorrespondingauthor}}
\ead{kaisa.laiho@dlr.de}
\cortext[mycorrespondingauthor]{Corresponding author}
\author[mpl,eruni]{T.~Dirmeier}
\author[mpl,eruni]{G. Shafiee}
\author[eruni,mpl]{and Ch.~Marquardt} 
\address[ulm]{German Aerospace Center (DLR e.V.), Institute of Quantum Technologies, Wilhelm-Runge-Str. 10, 89081 Ulm, Germany}
\address[mpl]{Max Planck Institute for the Science of Light, Staudtstr.~2, 91058 Erlangen, Germany}
\address[eruni]{Friedrich-Alexander-Universit\"at Erlangen-N\"urnberg (FAU), Department of Physics, Staudtstr.~7/B2, 91058 Erlangen, Germany}

\begin{abstract}
The Hong-Ou-Mandel (HOM) interference is one of the most intriguing quantum optical phenomena and crucial in performing quantum optical communication and computation tasks. Lately, twin beam emitters such as those relying on the process of parametric down-conversion (PDC) have become confident sources of heralded single photons.  
However, if the pump power is high enough, the pairs produced via PDC---often called signal and idler---incorporate multiphoton contributions that usually distort the investigated quantum features. Here, we derive the temporal characteristics of the HOM interference between heralded states from two independent narrowband PDC sources. Apart from the PDC multiphoton content, our treatment also takes into account effects arriving from an unbalanced beam splitter ratio and optical losses. We perform a simulation in the telecommunication wavelength range and provide a useful tool for finding the optimal choice for PDC process parameters. Our results offer insight in the properties of narrowband PDC sources and turn useful when driving quantum optical applications with them.
\end{abstract}

\end{frontmatter}


\section{Introduction}

The two-photon quantum interference named after the seminal experiment of Hong, Ou and Mandel provides evidence of photon bunching and stems from the true quantum features of light \cite{Hong1987}. Today, this quantum phenomenon is one of the most important building blocks of photonic quantum information and communication applications \cite{Bouchard2021}. Many quantum optical technologies require spectrally tailored photon sources in order to guarantee the applicability of the quantum light with the so-called quantum hardware. Hereby, modeling the investigated quantum optics task precisely and accurately with the required photonic source plays an important role \cite{Riedmatten2005, Schunk2016, Yu2022}.

Regarding photon-pair generation processes, the HOM interference between independent heralded single photons has vastly been investigated on the platforms relying both on the four-wave mixing \cite{Harada2011, Qian2016, Jeong2017, Davidson2021} and parametric down-conversion (PDC) \cite{Halder2007, Xue2010, Aboussouan2010, Hua2021}. Recently, highly versatile and conveniently tunable sources of photon pairs---usually denoted as signal and idler---have  successfully  been demonstrated with different spectro-temporal characteristic \cite{Halder2008, Foertsch2013, Mottola2020, Sultanov2022}. When utilizing photon-pair processes in the HOM interference experiments, care has to be taken since their multiphoton contributions are known to diminish the visibility of the two-photon interference dip. Such results have been reported at least when investigating the indistinguishability of signal and idler \cite{Cosme2008, Takeoka2015, Guenthner2015} and when examining the HOM interference between heralded states from different photon-pair sources \cite{Qian2016, Faruque2019} as well as between heralded single photons and coherent states \cite{Kolenderski2008, Laiho2009, Xu2023}. In case the multiphoton contributions are not taken into account the interpretation of the quantum features of the manipulated light can be falsified. In other cases, recently spectral multiplexing have been proposed and realized to counteract the drawback of PDC multiphoton contributions \cite{Hiemstra2020}.

The spectral and temporal characteristics of photons emitted via PDC are related to each other with the Fourier transform. Often, the spectral properties of these sources can be easily accessed \cite{Jin2015} and measured at the few photon level with high-resolution spectrographs \cite{Laiho2016}. Contrariwise, the coincidence discrimination mostly happens in the temporal domain. We take use of the spectral PDC properties to derive the temporal characteristics of the HOM interference dip between heralded states from two independent narrowband PDC sources. 
We emphasize that when it comes to time-resolved measurements, the treatment of the PDC sources including continuous-wave pumping \cite{Blauensteiner2009, Foertsch2013} and those including pulsed pumping \cite{Zukowski1995, Mosley2008, Tanida2012} often slightly differs from each other. While in the former case one can do truly time-resolved measurements of the quantum light, in the latter case one almost always needs to average over the duration of the pulse due to the slowness of the most optical detectors \cite{Christ2011}. In our treatment we assume an ideal time-resolving detection and therefore omit such averaging effects caused experimentally mostly by the photodection. Instead we scrutinize the effect of the PDC multiphoton contributions and typical imperfections in the optical arrangement such as optical losses and an unbalanced beam splitter ratio. Our derivation is based on evaluating the required expectation values directly with PDC states, in other words, two-mode squeezed vacuum states, which allows to easily capture the full state characteristics \cite{Iskhakov2013, Alsing2022}. Firstly, in our treatment there is no need of operating in the more cumbersome photon-number basis, which includes summation of photon numbers upto infinity or truncation of the sum, which might introduce artefacts. Secondly, we model the PDC process in the frequency space, which means that the spectral properties of signal and idler are intrinsically included in our calculation. For simplicity in our simulation we assume that the two PDC processes are identical. However, experimentally probably most challenging to realize is the spectral overlap of the individual PDC processes, especially when the heralded states are not created along a common path but rather via two independent crystals or waveguides \cite{Mosley2008, Tanida2012}.

We simulate the four-fold coincidence probability in the telecommunication wavelength range assuming Gaussian PDC spectra and indicate how the visibility of the HOM interference dip degrades in terms of the PDC multiphoton contributions. We note that each platform producing continuous-wave photon pairs has its highly individual spectral characteristics that  needs to be thoroughly scrutinized to gain an accurate estimate of the background effects \cite{Luo2015, Shafiee2020}. 
Further, we investigate the temporal characteristics of signal and idler cross-correlation \cite{Averchenko2020} to show the practicality of this figure-of-merit as a PDC process parameter since the integral over it turns out to be expedient for calibrating the mean photon flux of the source in loss-independent manner. We believe our results can enhance the efficient utilization of narrowband PDC sources in quantum optical applications and help reaching a high visibility in the HOM interference experiment.

\section{Derivation of the temporal characteristics of the HOM interference}

\begin{figure}[!tb]
\begin{center}
\includegraphics[width = 0.45\textwidth]{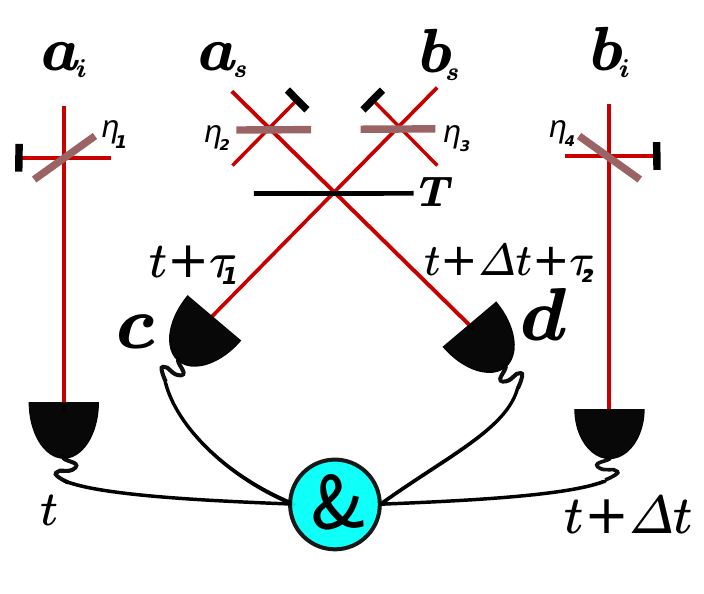}
\caption{Sketch of the HOM interference experiment with two heralded photons from different PDC sources. The idler ($i$) beams of the two narrowband PDC processes, which we label here with $a$ and $b$, are used as the heralds, while the signal ($s$) beams are sent to the input ports of the beam splitter having the transmittance $T$. Each beam path $a_{i}$, $a_{s}$, $b_{i}$ and  $b_{s}$ is objected to the optical loss, modeled as a beam splitter with the transmittance corresponding to the efficiency, at which the mode is detected. Vacuum modes are not explicitly denoted. For further details see text.}
\label{fig:sketch}
\end{center}
\end{figure}

We investigate the HOM interference between heralded photons from two independent narrowband PDC sources taking into account the PDC multiphoton contributions, optical losses at the paths of the signal and idler beams as well as an unbalanced beam splitter ratio. The optical arrangement is sketched in Fig.~\ref{fig:sketch} and the depicted four-fold coincidence probability can be derived via \cite{Qian2016}
\begin{eqnarray}
 P(t, t+\tau_{1}, \hspace{-1.8ex}&t+\Delta t +\tau_{2},&\hspace{-1.8ex} t+\Delta t) \nonumber \\
&=  \eta_{1}\eta_{4} & \hspace{-4ex} \braket{\hat{a}^{\dagger}_{i}(t) \hat{b}^{\dagger}_{i}(t+\Delta t)  \hat{c}^{\dagger}(t+\tau_{1}) \hat{d}^{\dagger}(t+\Delta t +\tau_{2}) \hat{d}(t+\Delta t+\tau_{2})  \hat{c}(t+\tau_{1})  \hat{b}_{i}(t+\Delta t) \hat{a}_{i}(t)},
\label{eq:P}
\end{eqnarray} 
in which $\hat{a}^{\dagger}_{\mu}$ [$\hat{a}_{\mu}$] and $\hat{b}^{\dagger}_{\mu}$ [$\hat{b}_{\mu}$] ($\mu = s,i$) stand for the photon creation [annihilation] in the signal ($s$) and idler ($i$) modes of the two  investigated PDC processes, $\eta_{\xi}$ ($\xi= 1\dots 4$) is the efficiency at the different optical paths (cf.~Fig.~\ref{fig:sketch}), and $\hat{c}^{\dagger}$ [$\hat{c}$] and $\hat{d}^{\dagger}$ [$\hat{d}$] stand for the photon creation [annihilation] operators at the beam splitter output ports. In Eq.~(\ref{eq:P}) the variables $t$, $t+\Delta t$ denote the times of detecting heralding events in modes $a$ and $b$, respectively, while $t+\tau_{1}$ and $t+\Delta t +\tau_{2}$ describe the times of detecting events in the detectors placed at the beam splitter output ports $c$ and $d$, respectively.

In order to evaluate Eq.~(\ref{eq:P}) we utilize the beam splitter transformations and trace over all vacuum modes. Thus, we replace the operators $\hat{c}$ and $\hat{d}$ at the beam splitter outputs with
\begin{eqnarray}
\hat{c} &\rightarrow &\sqrt{\eta_{2}(1-T)} \hat{a}_{s} +  \sqrt{\eta_{3}T} \hat{b}_{s} \quad  \nonumber \textrm{and}\\
\hat{d} &\rightarrow &\sqrt{\eta_{2}T} \hat{a}_{s} - \sqrt{\eta_{3}(1-T)} \hat{b}_{s},
\label{eq:BS}
\end{eqnarray}
in which  $T$ accounts for the used beam splitter transmittance,
whereafter we re-write Eq.~(\ref{eq:P}) in terms of the input modes $a$ and $b$. 
We model the PDC states as two-mode squeezed vacuum states, shortly \emph{twin beams}. After plugging Eq.~(\ref{eq:BS}) in Eq.~(\ref{eq:P}) we arrive at 16 terms, from which only six terms survive. The rest of the terms vanish, when the expectation values are evaluated with the twin beams.

The probability of measuring a four-fold coincidence click is then given by
\begin{eqnarray}
\label{eq:Prob}
P \hspace{-2ex}&(t, t+\tau_{1}, t+\Delta t +\tau_{2}, t+\Delta t) &  \\
 &\propto \eta_{1}\eta_{2}^{2}\eta_{4} T(1-T) & \hspace{-6ex}%
\braket{\hat{a}^{\dagger}_{i}(t)  \hat{a}^{\dagger}_{s}(t+\tau_{1}) \hat{a}^{\dagger}_{s}(t+\Delta t+\tau_{2})\hat{a}_{s}(t+\Delta t+\tau_{2})\hat{a}_{s}(t+\tau_{1}) \ \hat{a}_{i}(t) }\braket{\hat{b}^{\dagger}_{i}(t+\Delta t)\hat{b}_{i}(t+\Delta t)}  \nonumber \\
&+ \eta_{1}\eta_{3}^{2}\eta_{4} T(1-T)& \hspace{-6ex}%
\braket{\hat{a}^{\dagger}_{i}(t)\hat{a}_{i}(t)}\braket{\hat{b}^{\dagger}_{i}(t+\Delta t) \hat{b}^{\dagger}_{s}(t+\tau_{1})\hat{b}^{\dagger}_{s}(t+\Delta t+\tau_{2})\hat{b}_{s}(t+\Delta t+\tau_{2}) \hat{b}_{s}(t+\tau_{1}) \hat{b}_{i}(t+\Delta t)}  \nonumber \\
&\hspace{3ex}+ \eta_{1}\eta_{2}\eta_{3}\eta_{4} T^{2}&\hspace{-6ex}%
 \braket{\hat{a}^{\dagger}_{i}(t) \hat{a}^{\dagger}_{s}(t+\Delta t+\tau_{2})\hat{a}_{s}(t+\Delta t+\tau_{2}) \hat{a}_{i}(t)}\braket{\hat{b}^{\dagger}_{i}(t+\Delta t) \hat{b}^{\dagger}_{s}(t+\tau_{1})\hat{b}_{s}(t+\tau_{1}) \hat{b}_{i}(t+\Delta t)} \nonumber \\
&+ \eta_{1}\eta_{2}\eta_{3}\eta_{4} (1-T)^{2}& \hspace{-6ex}%
\braket{\hat{a}^{\dagger}_{i}(t) \hat{a}^{\dagger}_{s}(t+\tau_{1})\hat{a}_{s}(t+\tau_{1}) \hat{a}_{i}(t) }\braket{\hat{b}^{\dagger}_{i}(t+\Delta t)\hat{b}^{\dagger}_{s}(t+\Delta t+\tau_{2})\hat{b}_{s}(t+\Delta t+\tau_{2}) \hat{b}_{i}(t+\Delta t) } \nonumber \\
&- \eta_{1}\eta_{2}\eta_{3}\eta_{4} T(1-T)& \hspace{-5.5ex}%
 \braket{\hat{a}^{\dagger}_{i}(t) \hat{a}^{\dagger}_{s}(t+\tau_{1})\hat{a}_{s}(t+\Delta t+\tau_{2}) \hat{a}_{i}(t)}\braket{\hat{b}^{\dagger}_{i}(t+\Delta t) \hat{b}^{\dagger}_{s}(t+\Delta t+\tau_{2})\hat{b}_{s}(t+\tau_{1}) \hat{b}_{i}(t+\Delta t)} \nonumber \\
&- \eta_{1}\eta_{2}\eta_{3}\eta_{4} T(1-T)& \hspace{-5.5ex}%
\braket{\hat{a}^{\dagger}_{i}(t) \hat{a}^{\dagger}_{s} (t+\Delta t+\tau_{2})\hat{a}_{s}(t+\tau_{1})\hat{a}_{i}(t)}\braket{\hat{b}^{\dagger}_{i}(t+\Delta t) \hat{b}^{\dagger}_{s}(t+\tau_{1})\hat{b}_{s}(t+\Delta t+\tau_{2}) \hat{b}_{i}(t+\Delta t)}, \nonumber
\end{eqnarray}
in which the first two terms describe the effect of the PDC higher photon-number contributions to the coincidences, the two terms in the middle denote the contribution to the coincidences outside the HOM interference dip and the two last terms account for the two-photon quantum interference. 
We calculate the temporal properties of the HOM interference in terms of the temporal delay $\Delta t$ and define the visibility of the HOM dip as
$\mathcal{V} = \frac{P(\Delta t \rightarrow \infty) - P(\Delta t = 0) }{P(\Delta t \rightarrow \infty)}$.
The PDC multiphoton contributions are expected to produce an undesired background to the coincidences and thence, diminish the visibility of the investigated HOM interference dip.

Next, in order to be able to evaluate Eq.~(\ref{eq:Prob}) we introduce the characteristics of the twin beams. We note that for the sake of simplicity in the following we elaborate the required expectation values for the mode $a$ only. We emphasize that similar considerations apply for the mode $b$ also. The investigated states can be written in the form  $\ket{\Psi} = \hat{S}\ket{0}$, in which the narrowband squeezing operator takes the form \cite{Loudon2000}
\begin{eqnarray}
\hat{S} = \exp{\left( \int d \omega S^{*}(\omega_{s}) \hat{a}_{s}(\omega_{s})\hat{a}_{i}(\omega_{p}-\omega_{s})  -  \int d \omega S(\omega_{s}) \hat{a}_{s}^{\dagger}(\omega_{s})\hat{a}_{i}^{\dagger}(\omega_{p}-\omega_{s}) \right)},
\label{eq:cwSqueezer}
\end{eqnarray}
where $\omega_{s}$ describes the angular frequency of signal and $\omega_{p}$ is that of the pump beam. 
The joint spectral amplitude of signal and idler is given by $f(\omega_{s}, \omega_{i}) = S(\omega_{s})\delta(\omega_{p}-\omega_{s}-\omega_{i})$ with the complex function $S(\omega_{s}) = r(\omega_{s})\textrm{exp}[i\vartheta(\omega_{s})]$ having the amplitude $r(\omega_{s})$ and phase $\vartheta(\omega_{s})$. 
The twin beams obey the transformations \cite{Loudon2000, Barnett1997}
\begin{eqnarray}
\label{eq:Ss}
\hat{S}^{\dagger}\hat{a}_{s}(\omega_{s}) \hat{S} &= & \hat{a}_{s}(\omega_{s}) \beta(\omega_{s})- \hat{a}^{\dagger}_{i}(\omega_{p}-\omega_{s}) \alpha(\omega_{s}) \textrm{,} \\
\hat{S}^{\dagger}\hat{a}^{\dagger}_{s}(\omega_{s}) \hat{S} & = &\hat{a}^{\dagger}_{s}(\omega_{s}) \beta(\omega_{s})- \hat{a}_{i}(\omega_p-\omega_{s}) \alpha^{*}(\omega_{s}) \nonumber
\end{eqnarray}
and
\begin{eqnarray}
\label{eq:Si}
\hat{S}^{\dagger}\hat{a}_{i}(\omega_{p}-\omega_{s}) \hat{S} &=  &\hat{a}_{i}(\omega_p-\omega_{s}) \beta(\omega_{s})- \hat{a}^{\dagger}_{s}(\omega_{s}) \alpha(\omega_{s}), \\
\hat{S}^{\dagger}\hat{a}^{\dagger}_{i}(\omega_{p}-\omega_{s}) \hat{S} &=  &\hat{a}^{\dagger}_{i}(\omega_p-\omega_{s}) \beta(\omega_{s})- \hat{a}_{s}(\omega_{s}) \alpha^{*}(\omega_{s}) \nonumber
\end{eqnarray}
for signal and idler, respectively.
In Eqs~(\ref{eq:Ss}) and (\ref{eq:Si}) we utilize the short notations $\beta(\omega_{s}) =\textrm{cosh}[r(\omega_{s})]$ and $\alpha(\omega_{s}) = \textrm{sinh}[r(\omega_{s})]\textrm{exp}[i\vartheta(\omega_{s})]$, for which it applies $|\beta(\omega_{s})|^{2}-|\alpha(\omega_{s})|^{2}= 1$. 
We are interested in evaluating Eq.~(\ref{eq:Prob}) in the frequency space and take use of the Fourier transforms
\begin{eqnarray}
\hat{a}_{\mu}(t) &=& \frac{1}{\sqrt{2\pi}} \int d\omega \hat{a}_{\mu}(\omega_{\mu})\textrm{e}^{-i\omega_{\mu} t} \nonumber \quad \textrm{and} \\
\hat{a}^{\dagger}_{\mu}(t) &=& \frac{1}{\sqrt{2\pi}} \int d\omega \hat{a}^{\dagger}_{\mu}(\omega_{\mu})\textrm{e}^{i\omega_{\mu} t}
\label{eq:Fourier}
\end{eqnarray}
and finally, plug Eqs~(\ref{eq:Ss}) and (\ref{eq:Si}) together with Eq.~(\ref{eq:Fourier}) in Eq.~(\ref{eq:Prob}) order to evaluate it. 

Following the treatment in Ref.~\cite{Loudon2000} we determine the expectation values required for evaluating Eq.~(\ref{eq:Prob}). For the detailed derivation see \ref{sec:app}. The mean photon flux takes the form 
\begin{eqnarray}
\mathcal{N}= \braket{\hat{a}^{\dagger}_{i}(t)\hat{a}_{i}(t)}  = \braket{\hat{a}^{\dagger}_{s}(t)\hat{a}_{s}(t)}  = \frac{1}{2\pi}\int d\omega  \left | \alpha(\omega)\right |^{2},
\end{eqnarray}
and the coherence time of the source can be determined via the normalized first-order coherence $g^{(1)}(\tau) = \frac{\braket{\hat{a}^{\dagger}_{\mu}(t)\hat{a}_{\mu}(t+\tau)}}{\braket{\hat{a}^{\dagger}_{\mu}(t)\hat{a}_{\mu}(t)}} =  \frac{1}{2\pi \mathcal{N}}\int d\omega  \left | \alpha(\omega)\right |^{2} \textrm{e}^{-i\omega\tau}$, which delivers
\begin{eqnarray}
\label{eq:tc}
\tau_{c} = \int d\tau|g^{(1)}(\tau)|^{2} =  \frac{1}{2\pi \mathcal{N}^{2}} \hspace{-0.3ex} \int \hspace{-0.3ex} d\omega  |\alpha(\omega)|^{2} \hspace{-0.3ex}\int \hspace{-0.3ex} d\tilde{\omega}  |\alpha(\tilde{\omega})|^{2} \underbrace{ \frac{1}{2\pi}  \int d \tau  \  \textrm{e}^{i(\omega-\tilde{\omega})\tau}}_{= \delta(\omega-\tilde{\omega})} =  \frac{1}{2\pi \mathcal{N}^{2}} \hspace{-0.3ex} \int \hspace{-0.3ex} d\omega  |\alpha(\omega)|^{4},
\end{eqnarray}
and which can be measured in an interferometric experiment \cite{Blauensteiner2009}. 
Alternatively, the coherence time can also be determined via the second-order correlation of an individual marginal beam that is~signal or idler (c.f.~\ref{sec:app}).
Thus, experimentally it can be accessed in a Hanbury-Brown-Twiss experiment of the investigated marginal beam \cite{Brown1956, Shafiee2020}. Luckily, the time-integral over a correlation function is tolerant against the jitter of the detectors and do not disturb the evaluation \cite{Blauensteiner2009, Kreinberg2017,Laiho2022}.

Another important figure-of-merit, namely the signal-idler cross-correlation, can be re-written as
\begin{eqnarray}
&\braket{\hat{a}^{\dagger}_{i}(t^{\prime}) \hat{a}^{\dagger}_{s}(t)\hat{a}_{s}(t) \hat{a}_{i}(t^{\prime}) } =\left[ \frac{1}{2\pi} \hspace{-0.3ex}\int \hspace{-0.3ex}d\omega  \left | \alpha(\omega)\right |^{2}  \right]^{2} + \frac{1}{2\pi} \hspace{-0.3ex} \int \hspace{-0.3ex} d\omega  \alpha^{*}(\omega) \beta(\omega)\textrm{e}^{-i\omega( t^{\prime}-t)} \frac{1}{2\pi}\hspace{-0.3ex}\int \hspace{-0.3ex} d\tilde{\omega}  \alpha(\tilde{\omega}) \beta(\tilde{\omega})\textrm{e}^{i\tilde{\omega} (t^{\prime}-t)},
\label{eq:I}
\end{eqnarray}
which in the normalized form of 
\begin{eqnarray}
g^{(2)}_{s,i}(\tau = t-t^{\prime}) =\frac{ \braket{\hat{a}^{\dagger}_{i}(t^{\prime}) \hat{a}^{\dagger}_{s}(t)\hat{a}_{s}(t)\hat{a}_{i}(t^{\prime})} }{\braket{\hat{a}^{\dagger}_{s}(t)\hat{a}_{s}(t)} \braket{\hat{a}^{\dagger}_{i}(t^{\prime})\hat{a}_{i}(t^{\prime})}},
\label{eq:gsi}
\end{eqnarray}
delivers the important information of the coincidences-to-accidentals ratio of the PDC source denoted here with $g^{(2)}_{s,i}(\tau  = 0)$. Moreover, Eq.~(\ref{eq:gsi}) usually can rather conveniently be measured via the coincidence discrimination between signal and idler \cite{Foertsch2015}. Fortunately, the strength of the signal and idler cross-correlation offers means for the quantification of the PDC higher photon-number contributions and  can be used as a calibration tool in order to guarantee a high visibility in the HOM interference experiment.

Additionally, we note that the integral over the function $g^{(2)}_{s,i}(\tau) -1$ delivers a loss-independent estimate of the mean photon flux given as
\begin{eqnarray}
\int  d \tau \left [ g^{(2)}_{s,i}(\tau) -1 \right]& =&\frac{ \int d \tau \frac{1}{2\pi} \hspace{-0.3ex} \int \hspace{-0.3ex} d\omega  \alpha^{*}(\omega) \beta(\omega)\textrm{e}^{i\omega \tau} \frac{1}{2\pi}\hspace{-0.3ex}\int \hspace{-0.3ex} d\tilde{\omega}  \alpha(\tilde{\omega}) \beta(\tilde{\omega})\textrm{e}^{-i\tilde{\omega}\tau}}{\left [\frac{1}{2\pi} \hspace{-0.3ex}\int \hspace{-0.3ex}d\omega  \left | \alpha(\omega)\right |^{2}  \right]^{2}} \nonumber \\
& = & \frac{1}{2\pi \mathcal{N}^{2}} \hspace{-0.3ex} \int \hspace{-0.3ex} d\omega  \alpha^{*}(\omega) \beta(\omega)\hspace{-0.3ex}\int \hspace{-0.3ex} d\tilde{\omega}  \alpha(\tilde{\omega}) \beta(\tilde{\omega}) \underbrace{ \frac{1}{2\pi}  \int d \tau \   \textrm{e}^{i(\omega-\tilde{\omega})\tau}}_{= \delta(\omega-\tilde{\omega})} \nonumber \\
&=&\frac{1}{\mathcal{N}^{2}} \  \frac{1}{2\pi} \hspace{-0.3ex} \int \hspace{-0.3ex} d\omega \left | \alpha(\omega)\right |^{2}|\beta(\omega)|^{2} 
= \frac{1}{\mathcal{N}} +  \frac{1}{2\pi \mathcal{N}^{2}} \hspace{-0.3ex} \int \hspace{-0.3ex} d\omega \left | \alpha(\omega)\right |^{4} = \frac{1}{\mathcal{N}} + \tau_{c},
\label{eq:area}
\end{eqnarray}
in the last step of which we plugged in the outcome from Eq.~(\ref{eq:tc}). The result derived in Eq.~(\ref{eq:area}) is in analog to the case of the broadband multi-mode PDC, in which the time-integrated signal-idler cross-correlation delivers a sum of two terms, namely the inverse of the mean photon number and the inverse of the number of the excited modes \cite{Christ2011}. Moreover, we note that the time integral over a normalized correlation function have recently turned into a practical tool when carrying out quantum optical investigations of other photonic emitters also, such as nanolasers \cite{Kreinberg2017}. 
 
Further, for the evaluation of Eq.~(\ref{eq:Prob}) we require a more sophisticated signal-idler cross-correlation term given by
\begin{eqnarray}
\braket{\hat{a}^{\dagger}_{i}(t)\hat{a}^{\dagger}_{s}(t^{\prime})\hat{a}_{s}(t^{\prime\prime}) \hat{a}_{i}(t)} && \nonumber\\
&\hspace{-35ex} =& \hspace{-18ex} \left( \frac{1}{2\pi}\right)^{2} \hspace{0ex}\bigg[ \hspace{-0.3ex}\int \hspace{-0.3ex}d\omega | \alpha(\omega) |^{2}  \hspace{-0.3ex}\int \hspace{-0.3ex} d \tilde{\omega} \left | \alpha(\tilde{\omega})\right |^{2}  \textrm{e}^{i\tilde{\omega}(t^{\prime}-t^{\prime\prime})} + \hspace{-0.3ex} \int \hspace{-0.3ex} d\omega  \alpha^{*}(\omega) \beta(\omega)\textrm{e}^{-i\omega(t-t^{\prime})} 
\hspace{-0.3ex}\int \hspace{-0.3ex} d\tilde{\omega}  \alpha(\tilde{\omega}) \beta(\tilde{\omega})\textrm{e}^{i\tilde{\omega}(t-t^{\prime\prime})}
\bigg].
\label{eq:II}
\end{eqnarray}
Finally, in order to estimate the effect of the PDC multiphoton contributions to the coincidences within the HOM interference dip we take use of the expectation value
\begin{eqnarray}
&&\hspace{-5ex}\braket{\hat{a}^{\dagger}_{i}(t)\hat{a}^{\dagger}_{s}(t^{\prime})\hat{a}^{\dagger}_{s}(t^{\prime\prime})\hat{a}_{s}(t^{\prime\prime})\hat{a}_{s}(t^{\prime}) \hat{a}_{i}(t)  } \nonumber \\
&=& \hspace{-1ex}\left ( \frac{1}{2\pi} \right)^{3} \hspace{-0.3ex}\int \hspace{-1.5ex}\int \hspace{-0.3ex}    d\omega  d \tilde{\omega} d\omega^{\prime}  \
 \alpha(\omega) \beta(\omega)  \left|\alpha(\tilde{\omega} )\right|^{2} \alpha^{*}(\omega^{\prime}) \beta(\omega^{\prime}) 
\left[  \textrm{e}^{-i \omega (t^{\prime\prime}-t)} \textrm{e}^{i \tilde{\omega} (t^{\prime\prime}-t^{\prime})}   \textrm{e}^{i \omega^{\prime} (t^{\prime}-t)} +
 \textrm{e}^{-i \omega (t^{\prime}-t)}    \textrm{e}^{i \tilde{\omega} (t^{\prime}-t^{\prime\prime})} \textrm{e}^{i \omega^{\prime} (t^{\prime\prime}-t)} \right] 
\nonumber \\
&+& \hspace{-1ex}\left ( \frac{1}{2\pi}\right)^{3} \hspace{-0.3ex}\int \hspace{-1.5ex}\int \hspace{-0.3ex}   d\omega  d \tilde{\omega} d\omega^{\prime}  \
\alpha(\omega) \beta(\omega)  \left|\alpha(\tilde{\omega} )\right|^{2}  \alpha^{*}(\omega^{\prime}) \beta(\omega^{\prime}) 
\left[  \textrm{e}^{-i \omega (t^{\prime}-t)} \textrm{e}^{i \omega^{\prime} (t^{\prime}-t)} +
  \textrm{e}^{-i \omega (t^{\prime\prime}-t)}  \textrm{e}^{i \omega^{\prime} (t^{\prime\prime}-t)}\right]
 \nonumber \\
&+& \hspace{-1ex}\left ( \frac{1}{2\pi}\right)^{3} \hspace{-0.3ex}\int \hspace{-1.5ex}\int \hspace{-0.3ex}     d\omega  d \tilde{\omega} d\omega^{\prime}\
\left | \alpha(\omega)\right |^{2}  \left|\alpha(\tilde{\omega} )\right|^{2}   \left|\alpha(\omega^{\prime})\right|^{2}
\left[ 1+   \textrm{e}^{i \tilde{\omega} (t^{\prime}-t^{\prime\prime})} \textrm{e}^{i \omega^{\prime} (t^{\prime\prime}-t^{\prime})}  \right].
\label{eq:III}
\end{eqnarray}
\section{Simulation of the four-fold coincidences}

\begin{figure}[!b]
\begin{center}
\includegraphics[width = 0.87\textwidth]{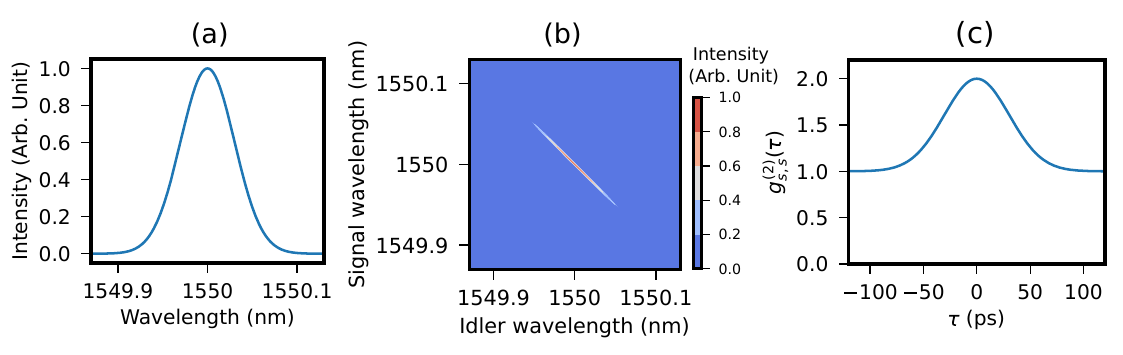}
\caption{(a) The marginal spectrum of the simulated narrowband PDC source. (b) The corresponding estimated joint spectral intensity of signal and idler. (c) The second-order correlation of an individual marginal beam, here $g^{(2)}_{s,s}(\tau)$.}
\label{fig:marginal_spectra}
\end{center}
\end{figure}

Next, we perform a simulation of the temporal characteristics of the HOM interference with narrowband PDC sources  in the telecommunication wavelength range. Our emphasis lies on understanding the effect of the multiphoton contributions on the HOM dip visibility. For the sake of simplicity we pressume that the spectral properties of the two deployed PDC sources are identical. Further, we assume a real-valued Gaussian spectral profile for the marginal beams, thus setting $\vartheta =0$. We express the spectral amplitude in the form \cite{Loudon2000}
\begin{eqnarray}
r(\omega_{s}) = \left ( \frac{2\pi \mathcal{F}^{2}}{\Delta^{2}}\right )^{1/4} \exp{\left (- \frac{[\omega_{p}/2-\omega_{s}]^{2}}{4\Delta^{2}}\right )},
\label{eq:S}
\end{eqnarray}
in which $\Delta$ is the spectral width of the marginal beams and the degeneracy lies at the angular frequency of $\omega_{p}/2$. In Eq.~(\ref{eq:S}) the value of $\mathcal{F}$ corresponds to the mean photon flux $\mathcal{N}$ in case the approximation $\sinh[r(\omega)]\approx r(\omega)$ holds. 

We plot the marginal spectrum in Fig.~\ref{fig:marginal_spectra}(a) having the degeneracy wavelength of \SI{1550}{\nano\meter} and a full-width-half-maximum (FWHM) of \SI{0.05}{\nano\meter}. Further, in Fig.~\ref{fig:marginal_spectra}(b) we illustrate the estimated joint spectral intensity of signal and idler, which as expected is strongly correlated in the frequency space. Moreover, in Fig.~\ref{fig:marginal_spectra}(c) we evaluate the second-order correlation of an individual marginal beam (here signal) denoted as $g^{(2)}_{s,s}(\tau)$, from which the coherence time of the PDC source can be estimated via the integration that delivers $\tau_{c} \approx \ $\SI{75.6}{\pico\second}  (c.f.~\ref{sec:app}).

\begin{figure}[!t]
\begin{center}
\includegraphics[width = 0.9\textwidth]{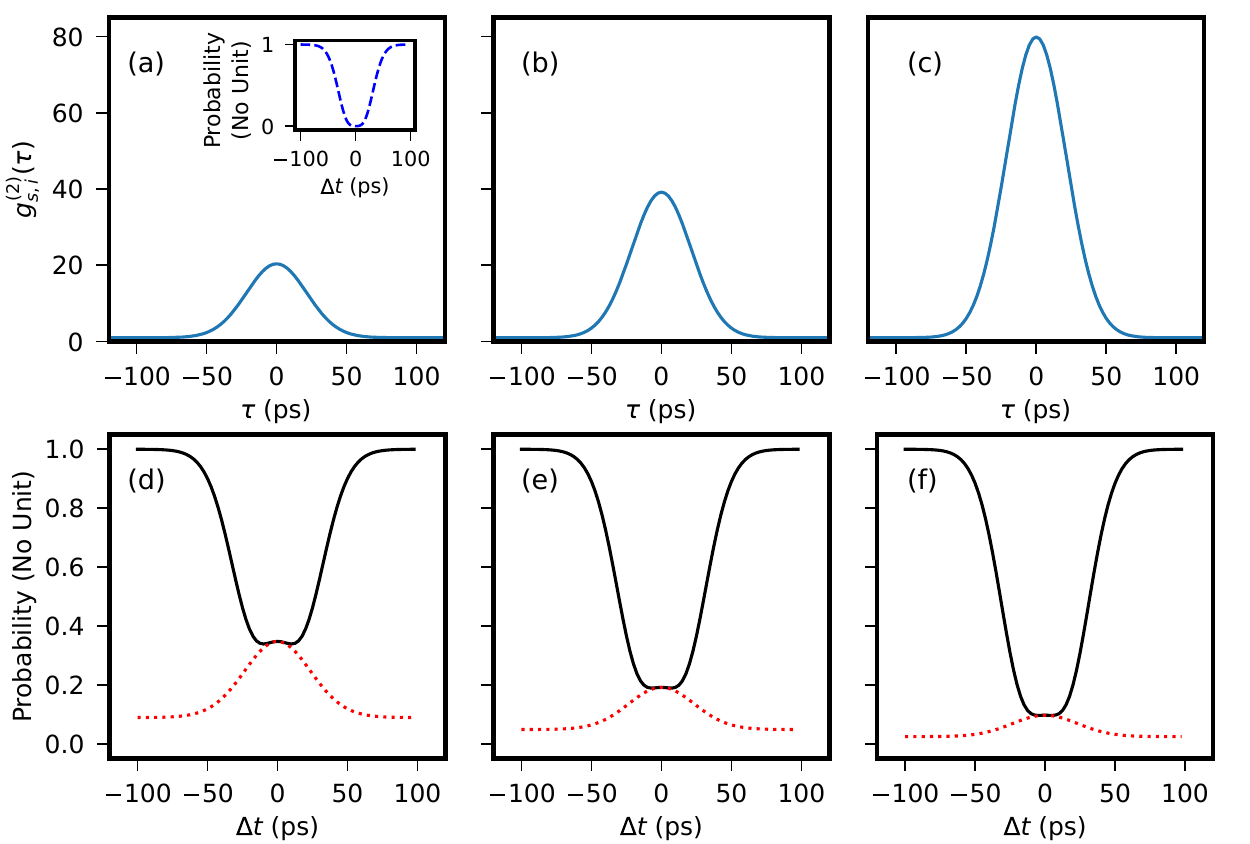}
\caption{The signal-idler correlation $g^{(2)}_{s,i}(\tau)$ taking the values of approximately (a) $20$ (b) $40$ and (c) $80$ at $\tau = 0$ together with the corresponding temporal characteristics of the HOM interference dips (black lines) resulting in the visibilities close to   (d) $0.65$ (e) $0.8$ and (f) $0.9$, respectively. The contribution to the coincidences  arriving from the multiphoton effects is plotted in (d-f) with red dotted lines. The inset in (a) illustrates the temporal properties of the HOM interference dip, if the multiphoton effects are neglected.}
\label{fig:gsi}
\end{center}
\end{figure}

We start by investigating the temporal characteristics of the HOM interference in the ideal case of having $T = 1/2$ and $\eta_{\xi= 1\dots4} = 1$. We evaluate the signal-idler cross-correlation at three different photon fluxes that result in $g^{(2)}_{s,i}(\tau =0)$-values of approximately $20$, $40$ and $80$ as shown in Figs~\ref{fig:gsi}(a-c), respectively. 
Next, we evaluate the temporal characteristics of the HOM interference dips with Eq.~(\ref{eq:Prob}) and present them in Figs~\ref{fig:gsi}(d-f) that yield the $\mathcal{V}$-values close to 0.65, 0.8 and 0.9, respectively.
Additionally, in Figs~\ref{fig:gsi}(d-f) we depict the contribution to the coincidences arriving from the multiphoton effects with red dotted lines. Clearly, their effect cannot be neglected and this background strongly diminishes the visibility of the HOM interference dips. As a consequence, a high value of the signal-idler cross-correlation $g^{(2)}_{s,i}(\tau =0)$ is required to counteract the effect of multiphoton contributions.
To this end, we note that in case the effect of the multiphoton contributions were neglected in Eq.~(\ref{eq:Prob}), we would expect almost a perfect HOM interference dip as illustrated in the inset in Fig.~\ref{fig:gsi}(a).

\begin{figure}[!t]
\begin{center}
\includegraphics[width = 0.9\textwidth]{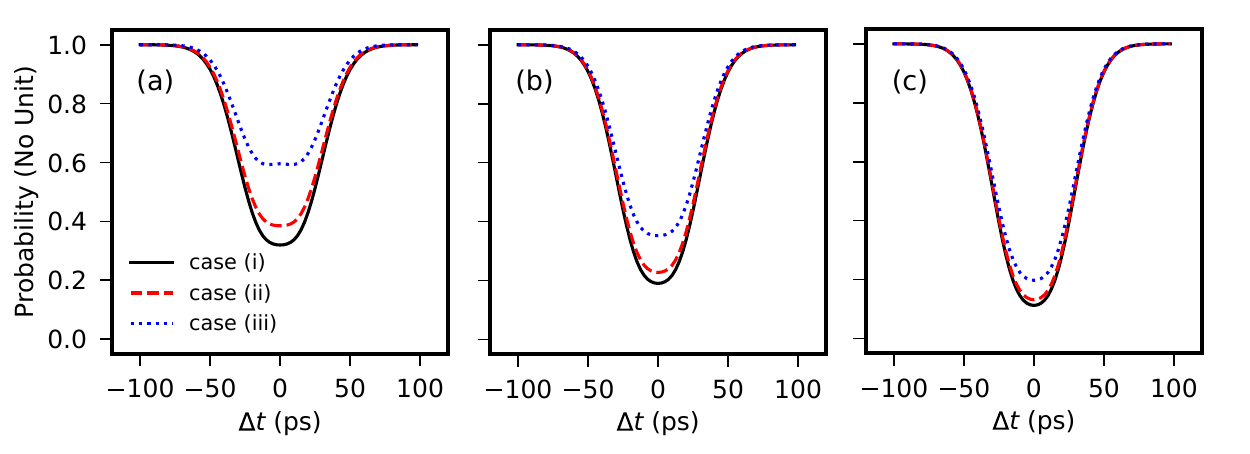}
\caption{Effect of experimental imperfections on the temporal characteristics of the HOM interference dips. We evaluate Eq.~(\ref{eq:Prob}) near the values of $g^{(2)}_{s,i}(\tau =0)$ of (a) $20$ (b) $40$ and (c) $80$ in the three investigated cases (i)-(iii). For the detailed description of the used parameters see the main text.}
\label{fig:losses}
\end{center}
\end{figure}

Next, we investigate the effect of experimental imperfections on the temporal characteristics of the HOM interference and take into account an unbalanced beam splitter ratio and optical losses. Looking at Eq.~(\ref{eq:Prob}) one can directly conclude that these imperfections contribute differently. The different optical losses in the signal beams' paths influence the scaling of the multiphoton contributions, while the utilization of an unbalanced beam splitter also affects the terms accounting for the HOM interference.
We investigate the effect of these imperfections in three different cases. In case (i) we assume that the transmittance of the beam splitter takes the value $T=0.45$, which is rather typical for integrated optics, while the optical losses in the signal beams' paths are equal. Thereafter, we keep the beam splitter transmittance unchanged at $T=0.45$ and assume slightly different efficiencies in the paths of the signal beams taking in case (ii) the values $\eta_{2} = 2\eta_{3}= 0.1$. In case (iii) we regard a stronger imbalance between the optical efficiencies in the signal beams' paths of $\eta_{2} = 4\eta_{3}= 0.2$. According to our experience values such as considered in cases (ii-iii) are rather typical for experimental arrangements  with photon-pair sources in the telecommunication wavelengths \cite{Guenthner2015, Chen2018}.
In Fig.~\ref{fig:losses} we plot for the three investigated cases the temporal properties of the HOM interference at the same values of $g^{(2)}_{s,i}(\tau= 0)$ of about $20$, $40$ and $80$ as in Fig.~\ref{fig:gsi}. We note that the utilization of an unbalanced beam splitter in case (i) causes slight changes in the shape of the HOM interference dips and results in the $\mathcal{V}$-values of approximately $0.68$, $0.81$ and $0.89$,   respectively. Due to the unbalanced beam splitter ratio the scaling of the multiphoton effects is slightly lower than in the ideal case presented in Fig.~\ref{fig:gsi}, which indeed can even result in a higher visibility.
The moderately different losses in the paths of the signal beams in case (ii) diminishes the $\mathcal{V}$-values close to  $0.61$, $0.77$ and $0.87$, respectively. However, a stronger imbalance in the optical losses of the signal beams' paths such as the one investigated in case (iii) causes a significant drop in the visibilities of the investigated HOM interference dips. In case (iii) the extracted $\mathcal{V}$-values drop near to $0.41$, $0.65$ and $0.80$, respectively. Evidently, the efficient alignment of the photon-pair setup plays a crucial role in order to achieve a high visibility in the HOM interference experiment. 

\begin{figure}[!t]
\begin{center}
\includegraphics[width = 0.43\textwidth]{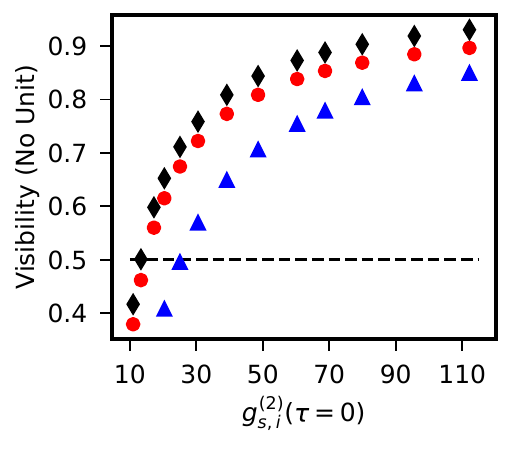}
\caption{Visibility of the HOM interference dip in terms of the value of $g^{(2)}_{s,i}(\tau=0)$. When regrading multiphoton contributions but excluding other experimental imperfections (black diamonds) the visibility of the HOM interference dip first grows fast in terms of  $g^{(2)}_{s,i}(\tau= 0)$ but then flattens out. When considering experimental imperfections  investigated in case (ii) (red circles), a slight deviation towards lower visibilities can be recognized, while the imperfections in case (iii)  (blue triangles) already result in a stronger deviation. The dashed horizontal line depicting the classical limit  of $\mathcal{V} =  0.5$ provides a guide for the eye.}
\label{fig:HOM}
\end{center}
\end{figure}

In order to gain insight, how the multiphoton contributions in the heralded states alter the characteristics of the HOM interference, we illustrate in Fig.~\ref{fig:HOM} the visibility of the HOM interference dip in terms of the value of the signal-idler cross-correlation at $\tau = 0$. When
regarding the ideal case (black diamonds) from Fig.~\ref{fig:gsi} that excludes experimental imperfections but takes into account multiphoton contributions we find that the classical limit of $\mathcal{V} = 0.5$, which applies for the coherent states of light, can in our case be rather easily met. For that purpose as low a value as $g^{(2)}_{s,i}(\tau= 0)\gtrsim 13$ is adequate. Interestingly, the HOM interference dip visibility grows fast with increasing signal-idler cross-correlation, but starts to flatten out near $g^{(2)}_{s,i}(\tau= 0)\approx 50$, where  $\mathcal{V} \approx 0.84$ can be reached. Beyond that value the visibility of the HOM interference dip grows only modestly. In order to reach the HOM interference dip visibility of $\mathcal{V} \approx  0.9$ the signal-idler cross-correlation needs to be increased to $g^{(2)}_{s,i}(\tau= 0)\approx80$. In order to suppress the multiphoton contributions, experimentally such high values of signal-idler cross-correlation can be achieved, however, meaning that the pump power  has to be low enough, which ultimately leads to lower count rates in the individual marginal beams \cite{Chen2018}. 
For comparison, we illustrate in Fig.~\ref{fig:HOM} the HOM interference dip visibilities also in cases (ii) (red circles) and (iii) (blue triangles). The regarded experimental imperfections do not change the observed behavior of the flattening, only the visibilities reached at the specific values of $g^{(2)}_{s,i}(\tau= 0)$ drop. Especially, if the efficiencies, at which the signal beams are detected, are strongly different different from each other such as in case (iii), it becomes difficult to reach a high visibility in the experiment.

\section{Conclusions}
We investigated the temporal characteristics of the HOM interference between heralded photons from two independent narrowband PDC sources. 
We derived equations for evaluating the temporal properties of the HOM interference dip, when taking into account the evident multiphoton contributions of PDC and other experimental imperfections such as optical losses and an unbalanced beam splitter ratio.
We performed a numerical simulation in the telecommunication wavelength range with a narrowband PDC source assuming a Gaussian spectral amplitude.
Our numerical simulation shows that the multiphoton contributions rather strongly diminish the visibility of the HOM interference dip.
Further, we find out that only modest changes are expected if the transmittance of the beam splitter slightly deviates from that of a perfectly balanced beam splitter as expected for realistic optics components.
More crucial is to take care that the optical losses in the signal beams' paths are comparable, since this affects most the scaling of the background from the multiphoton contributions. 
Strongly different losses on these beam paths results in lower visibilities of the HOM interference dips.
Additionally, we showed that the value of the normalized signal-idler cross-correlation $g^{(2)}_{s,i}(\tau=0)$, which corresponds to the coincidences-to-accidentals ratio, provides a useful tool for selecting a proper photon flux such that high visibility in the HOM interference experiment can be achieved.
Most interestingly, the time integral over the  function $g^{(2)}_{s,i}(\tau)-1$ that can usually be extracted in the experiment in a loss-independent manner is proportional to the inverse of the mean photon flux.
Altogether, our results show that the PDC multiphoton contributions have to be taken into consideration when investigating HOM interference with heralded PDC sources.
We deduce that our results are important for reaching high visibilities in the two-photon quantum interference experiments when utilizing narrowband photon-pair sources.  

\section{Acknowledgments}
We thank Alexander Otterpohl for helpful discussions.

\begin{thebibliography}{10}
\expandafter\ifx\csname url\endcsname\relax
  \def\url#1{\texttt{#1}}\fi
\expandafter\ifx\csname urlprefix\endcsname\relax\def\urlprefix{URL }\fi
\expandafter\ifx\csname href\endcsname\relax
  \def\href#1#2{#2} \def\path#1{#1}\fi

\bibitem{Hong1987}
C.~K. Hong, Z.~Y. Ou, L.~Mandel, Measurement of subpicosecond time intervals
  between two photons by interference, Phys.~Rev.~Lett. 59 (1987) 2044.

\bibitem{Bouchard2021}
F.~Bouchard, A.~Sit, Y.~Zhang, R.~Fickler, F.~M. Miatto, Y.~Yao, F.~Sciarrino,
  E.~Karimi, Two-photon interference: the hong–ou–mandel effect, Rep. Prog.
  Phys. 84 (2021) 012402.

\bibitem{Riedmatten2005}
H.~de~Riedmatten, I.~Marcikic, J.~A.~W. van Houwelingen, W.~Tittel, H.~Zbinden,
  N.~Gisin, Long-distance entanglement swapping with photons from separated
  sources, Phys. Rev. A 71 (2005) 050302(R).

\bibitem{Schunk2016}
G.~Schunk, U.~Vogl, F.~Sedlmeir, D.~V. Strekalov, A.~Otterpohl, V.~Averchenko,
  H.~G.~L. Schwefel, G.~Leuchs, Ch.~Marquardt, Frequency tuning of single
  photons from a whispering-gallery mode resonator to mhz-wide transitions, J.
  Mod. Optics 63 (2016) 2058.

\bibitem{Yu2022}
H.~Yu, C.~Yuan, R.~Zhang, Z.~Zhang, H.~Li, Y.~Wang, G.~Deng, L.~You, H.~Song,
  Z.~Wang, G.-C. Guo, Q.~Zhou, Spectrally multiplexed indistinguishable
  single-photon generation at telecom-band, Photonics Research 10 (2022) 1417.

\bibitem{Harada2011}
K.~Harada, H.~Takesue, H.~Fukuda, T.~Tsuchizawa, T.~Watanabe, K.~Yamada,
  Y.~Tokura, S.~Itabashi, Indistinguishable photon pair generation using two
  independent silicon wire waveguides, New J. Phys. 13 (2011) 065005.

\bibitem{Qian2016}
P.~Qian, Z.~Gu, R.~Cao, R.~Wen, Z.~Y. Ou, J.~F. Chen, W.~Zhang, Temporal purity
  and quantum interference of single photons from two independent cold atomic
  ensembles, Phys. Rev. Lett. 117 (2016) 013602.

\bibitem{Jeong2017}
T.~Jeong, Y.-S. Lee, J.~Park, H.~Kim, H.~S. Moon, Quantum interference between
  autonomous single-photon sources from doppler-broadened atomic ensembles,
  Optica 4 (2017) 1167.

\bibitem{Davidson2021}
O.~Davidson, R.~Finkelstein, E.~Poem, O.~Firstenberg, Bright multiplexed source
  of indistinguishable single photons with tunable ghz-bandwidth at room
  temperature, New J. Phys. 23 (2021) 073050.

\bibitem{Halder2007}
M.~Halder, A.~Beveratos, N.~Gisin, V.~Scarani, C.~Simon, H.~Zbinden, Entangling
  independent photons by time measurement, Nature Phys. 3 (2007) 692.

\bibitem{Xue2010}
Y.~Xue, A.~Yoshizawa, H.~Tsuchida, Hong-ou-mandel dip measurements of
  polarization-entangled photon pairs at 1550 nm, Optics Express 18 (2010)
  8182.

\bibitem{Aboussouan2010}
P.~Aboussouan, O.~Alibart, D.~B. Ostrowsky, P.~Baldi, S.~Tanzilli,
  High-visibility two-photon interference at a telecom wavelength using
  picosecond-regime separated sources, Phys. Rev. A 81 (2010) 021801(R).

\bibitem{Hua2021}
X.~Hua, T.~Lunghi, F.~Doutre, P.~Vergyris, G.~Sauder, P.~Charlier, L.~Labonté,
  V.~D’Auria, A.~Martin, S.~Tascu, M.~P.~D. Micheli, S.~Tanzilli, O.~Alibart,
  Configurable heralded two-photon fock-states on a chip, Opt. Express 29
  (2021) 415.

\bibitem{Halder2008}
M.~Halder, A.~Beveratos, R.~T. Thew, C.~Jorel, H.~Zbinden, N.~Gisin, High
  coherence photon pair source for quantum communication, New J. Phys. 10
  (2008) 023027.

\bibitem{Foertsch2013}
M.~Förtsch, J.~U. Fürst, C.~Wittmann, D.~Strekalov, A.~Aiello, M.~V.
  Chekhova, C.~Silberhorn, G.~Leuchs, Ch.~Marquardt, A versatile source of
  single photons for quantum information processing, Nature Comm. 4 (2013)
  1818.

\bibitem{Mottola2020}
R.~Mottola, G.~Buser, C.~M\"uller, T.~Kroh, A.~Ahlrichs, S.~Ramelow, O.~Benson,
  P.~Treutlein, J.~Wolters, An efficient, tunable, and robust source of
  narrow-band photon pairs at the 87rb d1 line, Opt. Express 28 (2020) 3159.

\bibitem{Sultanov2022}
V.~Sultanov, T.~Santiago-Cruz, M.~V. Chekhova, Flat-optics generation of
  broadband photon pairs with tunable polarization entanglement, Opt. Lett. 47
  (2022) 3872.

\bibitem{Cosme2008}
O.~Cosme, S.~Pádua, F.~A. Bovino, A.~Mazzei, F.~Sciarrino, F.~D. Martini,
  Hong-ou-mandel interferometer with one and two photon pairs, Phys. Rev. A 77
  (2008) 053822.

\bibitem{Takeoka2015}
M.~Takeoka, R.-B. Jin, M.~Sasaki, Full analysis of multi-photon pair effects in
  spontaneous parametric down conversion based photonic quantum information
  processing, New J. Phys. 17 (2015) 043030.

\bibitem{Guenthner2015}
T.~G\"unthner, B.~Pressl, K.~Laiho, J.~Geßler, S.~Höfling, M.~Kamp,
  C.~Schneider, G.~Weihs, Broadband indistinguishability from bright parametric
  downconversion in a semiconductor waveguide, J. Opt. 17 (2015) 125201.

\bibitem{Faruque2019}
I.~I. Faruque, G.~F. Sinclair, D.~Bonneau, T.~Ono, C.~Silberhorn, M.~G.
  Thompson, J.~G. Rarity, Estimating the indistinguishability of heralded
  single photons using second-order correlation, Phys. Rev. Applied 12 (2019)
  054029.

\bibitem{Kolenderski2008}
P. Kolenderski, K. Banaszek, Testing single-photon wave packets by Hong-Ou-Mandel 
interference, in the Proceedings of 2008 6th National Conference on Telecommunication 
Technologies and 2008 2nd Malaysia Conference on Photonics, (2008) pp. 94-97.
doi: 10.1109/NCTT.2008.4814246.

\bibitem{Laiho2009}
K.~Laiho, K.~N. Cassemiro, C.~Silberhorn, Producing high fidelity single
  photons with optimal brightness via waveguided parametric down-conversion,
  Opt.~Express 17 (2009) 22823.

\bibitem{Xu2023}
A.~Xu, L.~Duan, L.~Wang Y.~Zhang, Characterization of two-photon interference between a 
weak coherent state and a heralded single photon state,   Opt.~Express 31 (2023) 5662. 

\bibitem{Hiemstra2020}
T.~Hiemstra, T.~Parker, P.~Humphreys, J.~Tiedau, M.~Beck, M.~Karpiński,
  B.~Smith, A.~Eckstein, W.~Kolthammer, I.~Walmsley, Pure single photons from
  scalable frequency multiplexing, Phys. Rev. Applied 14 (2020) 014052.

\bibitem{Jin2015}
R.-B. Jin, T.~Gerrits, M.~Fujiwara, R.~Wakabayashi, T.~Yamashita, S.~Miki,
  H.~Terai, R.~Shimizu, M.~Takeoka, M.~Sasaki, Spectrally resolved
  hong-ou-mandel interference between independent photon sources, Optics
  Express 23 (2015) 28836.

\bibitem{Laiho2016}
K.~Laiho, B.~Pressl, A.~Schlager, H.~Suchomel, M.~Kamp, S.~H\"ofling,
  C.~Schneider, G.~Weihs, Uncovering dispersion properties in semiconductor
  waveguides to study photon-pair generation, Nanotechnology 27 (2016) 434003.

\bibitem{Blauensteiner2009}
B.~Blauensteiner, I.~Herbauts, S.~Betelli, A.~Poppe, H.~Hübel, Photon bunching
  in parametric down-conversion with continuous-wave excitation, Phys. Rev. A
  79 (2009) 063846.

\bibitem{Zukowski1995}
M. Zukowski, A. Zeilinger, H. Weinfurter, Entangling independent pulsed 
photon sources, Ann. N.Y. Acad. Sci. 755 (1995) 91.

\bibitem{Mosley2008}
P. J. Mosley, J. S. Lundeen, B. J. Smith, P. Wasylczyk, A. B. U’Ren, 
C. Silberhorn, I. A. Walmsley, Heralded Generation of Ultrafast Single Photons 
in Pure Quantum States, Phys. Rev. Lett. 100 (2008) 133601.

\bibitem{Tanida2012}
M. Tanida, R. Okamoto, S. Takeuchi, Highly indistinguishable heralded single-photon 
sources using parametric down conversion, Opt. Express 20 (2012) 15275.

\bibitem{Christ2011}
A.~Christ, K.~Laiho, A.~Eckstein, K.~N. Cassemiro, C.~Silberhorn, Probing
  multimode squeezing with correlation functions, New. J. Phys. 13 (2011)
  033027.

\bibitem{Iskhakov2013}
T.~S. Iskhakov, K.~Y. Spasibko, M.~V. Chekhova, G.~Leuchs, Macroscopic
  hong–ou–mandel interference, New J. Phys. 15 (2013) 093036.

\bibitem{Alsing2022}
P.~M. Alsing, R.~J. Birrittella, C.~C. Gerry, J.~Mimih, P.~L. Knight,
  Extending the hong-ou-mandel effect: The power of nonclassicality, Phys. Rev.
  A 105 (2022) 013712.

\bibitem{Luo2015}
K.~H. Luo, H.~Herrmann, S.~Krapick, B.~Brecht, R.~Ricken, V.~Quiring, H.~Suche,
  W.~Sohler, C.~Silberhorns, Direct generation of genuine
  single-longitudinal-mode narrowband photon pairs, New J. Phys. 17 (2015)
  073039.

\bibitem{Shafiee2020}
G.~Shafiee, D.~V. Strekalov, A.~Otterpohl, F.~Sedlmeir, G.~Schunk, U.~Vogl,
  H.~G.~L. Schwefel, G.~Leuchs, Ch.~Marquardt, Nonlinear power dependence of the
  spectral properties of an optical parametric oscillator below threshold in
  the quantum regime, New J. Phys. 22 (2020) 073045.

\bibitem{Averchenko2020}
V.~Averchenko, D.~Sych, Ch.~Marquardt, G.~Leuchs, Efficient generation of
  temporally shaped photons using nonlocal spectral filtering, Phys. Rev. A 101
  (2020) 013808.

\bibitem{Loudon2000}
R.~Loudon, The quantum theory of light, 3rd Edition, Oxford university Press,
  2000.

\bibitem{Barnett1997}
S.~M. Barnett, P.~M. Randmore, Methods in theoretical quantum optics, Oxford
  University Press, 1997.

\bibitem{Brown1956}
R.~Hanbury Brown, R.~Q. Twiss, A test of a new type of stellar interferometer on
  sirius, Nature 178 (1956) 1046.

\bibitem{Kreinberg2017}
S.~Kreinberg, W.~W. Chow, J.~W.~C. Schneider, C.~Gies, F.~Jahnke, S.~Höfling,
  M.~Kamp, S.~Reitzenstein, Emission from quantum-dot high-$\beta$
  microcavities: transition from spontaneous emission to lasing and the effects
  of superradiant emitter coupling, Light: Science \& Applications 6 (2017)
  17030.

\bibitem{Laiho2022}
K.~Laiho, T.~Dirmeier, M.~Schmidt, S.~Reitzenstein, C.~Marquardt, Measuring
  higher-order photon correlations of faint quantum light: a short review,
  Phys. Lett. A 435 (2022) 128059.

\bibitem{Foertsch2015}
M.~Förtsch, G.~Schunk, J.~U. Fürst, D.~Strekalov, T.~Gerrits, M.~J. Stevens,
  F.~Sedlmeir, H.~G.~L. Schwefel, S.~W. Nam, G.~Leuchs, Ch.~Marquardt, Highly
  efficient generation of single-mode photon pairs from a crystalline
  whispering-gallery-mode resonator source, Phys. Rev. A 91 (2015) 023812.



\bibitem{Klyshko1977}
D. N. Klyshko, Utilization of vacuum fluctuations as an optical brightness standard, 
Sov. J. Quantum Electron., 7 (1977) 591.

\bibitem{Harder2013}
G. Harder, V. Ansari, B. Brecht, T. Dirmeier, C. Marquardt, C. Silberhorn, 
An optimized photon pair source for quantum circuits, Opt. Express 21 (2013) 13975.

\bibitem{Krapick2013}
S. Krapick, H. Herrmann, V. Quiring, B. Brecht, H. Suche and C. Silberhorn, 
An efficient integrated two-color source for heralded single photons, New J. Phys. 
15 (2013) 033010.

\bibitem{Chen2018}
H.~Chen, S.~Auchter, M.~Prilm\"uller, A.~Schlager, T.~Kauten, K.~Laiho,
  B.~Pressl, H.~Suchomel, M.~Kamp, S.~H\"ofling, C.~Schneider, G.~Weihs,
  Invited article: Time-bin entangled photon pairs from bragg-reflection
  waveguides, APL Photonics 3 (2018) 080804.

\end{thebibliography}

\appendix
\section{\label{sec:app} Derivation of the expectation values}
In this appendix we derive for the completeness the expectation values required for evaluating Eq.~(\ref{eq:Prob}). We follow the treatment in Ref.~\cite{Loudon2000}, employ the Fourier transforms in Eq.~(\ref{eq:Fourier}) to replace the time domain with the frequency space and utilize the transformations in Eqs~(\ref{eq:Ss}) and (\ref{eq:Si}). We start by evaluating
\begin{eqnarray}
\braket{\hat{a}^{\dagger}_{s}(t)\hat{a}_{s}(t+\tau)}  \hspace{-1ex}&=&\hspace{-1ex} \frac{1}{2\pi}\int \hspace{-1.5ex} \int d\omega   d\tilde{\omega}\
\textrm{e}^{i\omega t} \textrm{e}^{-i\tilde{\omega} (t+\tau)} \braket{0|\hat{S}^{\dagger}
\hat{a}^{\dagger}_{s}(\omega)\hat{a}_{s}(\tilde{\omega})\hat{S}|0}\nonumber \\
 &=& \hspace{-1ex}\frac{1}{2\pi}\int \hspace{-1.5ex}\int d\omega   d\tilde{\omega}\
\textrm{e}^{i\omega t} \textrm{e}^{-i\tilde{\omega} (t+\tau)} \alpha^{*}(\omega)\alpha(\tilde{\omega}) \underbrace {\braket{\hat{a}_{i}(\omega_{p}-\omega)\hat{a}^{\dagger}_{i}(\omega_{p}-\tilde{\omega})}}_{= \delta(\omega -\tilde{\omega})}\nonumber \\
 &=& \hspace{-1ex} \frac{1}{2\pi}\int \hspace{-1.5ex}\int d\omega  \left | \alpha(\omega)\right |^{2} \textrm{e}^{-i\omega \tau},
\end{eqnarray}
in which we used the relation $\hat{S}\hat{S}^{\dagger} = 1$ and the commutator $[\hat{a}^{\dagger}_{\mu}(\omega), \hat{a}_{\mu}(\tilde{\omega})]= \delta(\omega-\tilde{\omega})$.
Next, we investigate the second-order correlation of an individual marginal beam (here signal) that takes the form
\begin{eqnarray}
\braket{\hat{a}^{\dagger}_{s}(t)\hat{a}^{\dagger}_{s}(t^{\prime}) \hat{a}_{s}(t^{\prime})\hat{a}_{s}(t)} \hspace{-1ex}
&=& \hspace{-1ex} \left(\frac{1}{2\pi}\right)^{2} \int \hspace{-1.5ex}\int d\omega_{1} d\omega_{2}  d\omega_{3}  d\omega_{4} \ \textrm{e}^{i\omega_{1} t} \textrm{e}^{i\omega_{2} t^{\prime}}
 \textrm{e}^{-i\omega_{3} t^{\prime}} \textrm{e}^{-i\omega_{4} t}\braket{ 0|\hat{S}^{\dagger}\hat{a}^{\dagger}_{s}(\omega_{1})\hat{a}^{\dagger}_{s}(\omega_{2}) \hat{a}_{s}(\omega_{3}) \hat{a}_{s}(\omega_{4})\hat{S}|0} \nonumber \\
&=& \hspace{-1ex} \left(\frac{1}{2\pi}\right)^{2}\int \hspace{-1.5ex} \int d\omega_{1} d\omega_{2}  d\omega_{3}  d\omega_{4} \ \textrm{e}^{i\omega_{1} t} \textrm{e}^{i\omega_{2} t^{\prime}}
 \textrm{e}^{-i\omega_{3} t^{\prime}} \textrm{e}^{-i\omega_{4} t}\alpha^{*}(\omega_{1})\alpha^{*}(\omega_{2})\alpha(\omega_{3})\alpha(\omega_{4}) \nonumber \\
&&\hspace{20ex} \underbrace{\braket{\hat{a}^{\dagger}_{i}(\omega_{p}-\omega_{1})\hat{a}^{\dagger}_{i}(\omega_{p}-\omega_{2}) \hat{a}_{i}(\omega_{p}-\omega_{3}) \hat{a}_{i}(\omega_{p}-\omega_{4})}}_{=\delta(\omega_{2}-\omega_{3})\delta(\omega_{1}-\omega_{4}) +\delta(\omega_{1}-\omega_{3})\delta(\omega_{2}-\omega_{4})},
\end{eqnarray}
which we re-write as
\begin{eqnarray}
\braket{\hat{a}^{\dagger}_{s}(t)\hat{a}^{\dagger}_{s}(t^{\prime}) \hat{a}_{s}(t^{\prime})\hat{a}_{s}(t)} 
\hspace{-1ex} &=& \hspace{-1ex} \left(\frac{1}{2\pi}\right)^{2} \left [\int d\omega  \left | \alpha(\omega)\right |^{2} \right]^{2} +\frac{1}{2\pi}\int d\omega  \left | \alpha(\omega)\right |^{2} \textrm{e}^{i\omega (t-t^{\prime})}\frac{1}{2\pi}\int d\tilde{\omega}  \left | \alpha(\tilde{\omega})\right |^{2} \textrm{e}^{-i\tilde{\omega} (t-t^{\prime})}.
\end{eqnarray}
Thence, the normalized form of the second-order correlation of an individual marginal beam delivers
\begin{eqnarray}
g^{(2)}_{s,s}(\tau = t-t^{\prime}) \hspace{-1ex}&=&\hspace{-1ex}\frac{ \braket{\hat{a}^{\dagger}_{s}(t) \hat{a}^{\dagger}_{s}(t^{\prime})\hat{a}_{s}(t^{\prime})\hat{a}_{s}(t)} }{\braket{\hat{a}^{\dagger}_{s}(t)\hat{a}_{s}(t)} \braket{\hat{a}^{\dagger}_{s}(t^{\prime})\hat{a}_{s}(t^{\prime})}} \nonumber \\
&=& \hspace{-1ex} 1 +\frac{1}{2\pi\mathcal{N}}\int d\omega  \left | \alpha(\omega)\right |^{2} \textrm{e}^{i\omega\tau
}\frac{1}{2\pi\mathcal{N}}\int d\tilde{\omega}  \left | \alpha(\tilde{\omega})\right |^{2} \textrm{e}^{-i\tilde{\omega}\tau},
\end{eqnarray}
which can also be used for evaluating the coherence time via $\tau_{c} =  \int d\tau \ \left ( g^{(2)}_{s,s}(\tau) -1\right )$.

Further, we determine the cross-correlation term
\begin{eqnarray}
\label{eq:appCross}
\braket{\hat{a}^{\dagger}_{i}(t)\hat{a}^{\dagger}_{s}(t^{\prime}) \hat{a}_{s}(t^{\prime \prime})\hat{a}_{i}(t)} &&\\
&\hspace{-20ex}= & \hspace{-10ex} \left(\frac{1}{2\pi}\right)^{2} \int \hspace{-1.5ex}\int d\omega_{1} d\omega_{2}  d\omega_{3}  d\omega_{4} \ \textrm{e}^{i\omega_{1} t} \textrm{e}^{i\omega_{2} t^{\prime}}
 \textrm{e}^{-i\omega_{3} t^{\prime\prime}} \textrm{e}^{-i\omega_{4} t}
\braket{0|\hat{S}^{\dagger}\hat{a}^{\dagger}_{i}(\omega_{1})\hat{a}^{\dagger}_{s}(\omega_{2}) \hat{a}_{s}(\omega_{3}) \hat{a}_{i}(\omega_{4})\hat{S}|0} \nonumber
\end{eqnarray}
which requires the expectation value
\begin{eqnarray}
\braket{\hat{a}^{\dagger}_{i}(\omega_{1})\hat{a}^{\dagger}_{s}(\omega_{2}) \hat{a}_{s}(\omega_{3}) \hat{a}_{i}(\omega_{4})} \nonumber \\
&\hspace{-25ex}=& \hspace{-13ex}\big< \hat{a}_{s}(\omega_{p}-\omega_{1})  \alpha^{*}(\omega_{p}-\omega_{1}) \big [ \hat{a}^{\dagger}_{s}(\omega_{2})\beta(\omega_{2}) - \hat{a}_{i}(\omega_{p}-\omega_{2})\alpha^{*}(\omega_{2})\big] \nonumber  \\
&\hspace{-25ex}& \hspace{-14ex} \otimes\big [\hat{a}_{s}(\omega_{3})\beta(\omega_{3}) - \hat{a}^{\dagger}_{i}(\omega_{p}-\omega_{3})\alpha(\omega_{3})\big]  \hat{a}^{\dagger}_{s}(\omega_{p}-\omega_{4})  \alpha(\omega_{p}-\omega_{4})\big> \\
&\hspace{-25ex}=& \hspace{-13ex}\alpha^{*}(\omega_{p}-\omega_{1}) \alpha(\omega_{p}-\omega_{4})\beta(\omega_{2}) \beta(\omega_{3}) \underbrace{\braket{ \hat{a}_{s}(\omega_{p}-\omega_{1})\hat{a}^{\dagger}_{s}(\omega_{2}) \hat{a}_{s}(\omega_{3})\hat{a}^{\dagger}_{s}(\omega_{p}-\omega_{4})}}_{=\delta(\omega_{p}-\omega_{1}-\omega_{2})\delta(\omega_{p}-\omega_{4}-\omega_{3})} \nonumber \\
&\hspace{-25ex}+&\hspace{-13ex}
 \alpha^{*}(\omega_{p}-\omega_{1}) \alpha(\omega_{p}-\omega_{4})\alpha^{*}(\omega_{2}) \alpha(\omega_{3}) \underbrace{ \braket{\hat{a}_{s}(\omega_{p}-\omega_{1})\hat{a}^{\dagger}_{i}(\omega_{p}-\omega_{2}) \hat{a}_{i}(\omega_{p}-\omega_{3})\hat{a}^{\dagger}_{s}(\omega_{p}-\omega_{4})}}_{=\delta(\omega_{1}-\omega_{4})\delta(\omega_{2}-\omega_{3})}.\nonumber 
\end{eqnarray}
Eventually, Eq.~(\ref{eq:appCross}) takes the form
\begin{eqnarray}
\braket{\hat{a}^{\dagger}_{i}(t)\hat{a}^{\dagger}_{s}(t^{\prime}) \hat{a}_{s}(t^{\prime \prime})\hat{a}_{i}(t)} 
\hspace{-1ex}&=& \hspace{-1ex}\frac{1}{2\pi} \int d\omega \alpha^{*}(\omega)\beta(\omega)  \textrm{e}^{-i\omega(t- t^{\prime})} \frac{1}{2\pi}  \int d\tilde{\omega} \alpha(\tilde{\omega})\beta(\tilde{\omega})\textrm{e}^{i\omega(t- t^{\prime\prime})} \nonumber \\
&+&\hspace{-1ex} \frac{1}{2\pi} \int d\omega |\alpha(\omega_{p}-\omega)|^{2} \frac{1}{2\pi}  \int 
d\tilde{\omega} |\alpha(\tilde{\omega})|^{2}\textrm{e}^{i\tilde{\omega}(t^{\prime}- t^{\prime\prime})},
\end{eqnarray}
which further reduces to Eq.~(\ref{eq:I}) if $t^{\prime} = t^{\prime\prime}$.

Finally, we evaluate the effect of the multiphoton contributions to the coincidences, which can be extracted via
\begin{eqnarray}
\label{eq:thpnc}
&&\hspace{-4ex}\braket{\hat{a}^{\dagger}_{i}(t)\hat{a}^{\dagger}_{s}(t^{\prime})\hat{a}^{\dagger}_{s}(t^{\prime\prime})\hat{a}_{s}(t^{\prime \prime})\hat{a}_{s}(t^{\prime}) \hat{a}_{i}(t)  } \\
&\hspace{-3.5ex}=& \hspace{-2.5ex}\left(\frac{1}{2\pi}\right)^{3} \hspace{-1ex}\int \hspace{-1.5ex} \int \hspace{-1ex} d\omega_{1} d\omega_{2}  d\omega_{3}  d\omega_{4}   d\omega_{5}  d\omega_{6} \ \textrm{e}^{i\omega_{1} t} \textrm{e}^{i\omega_{2} t^{\prime}}
 \textrm{e}^{i\omega_{3} t^{\prime\prime}} \textrm{e}^{-i\omega_{4} t^{\prime\prime}} \textrm{e}^{-i\omega_{5}t^{\prime}} \textrm{e}^{-i\omega_{6} t}
\braket{0|\hat{S}^{\dagger} \hat{a}^{\dagger}_{i}(\omega_{1})\hat{a}^{\dagger}_{s}(\omega_{2}) \hat{a}^{\dagger}_{s}(\omega_{3}) \hat{a}_{s}(\omega_{4}) \hat{a}_{s}(\omega_{5})  \hat{a}_{i}(\omega_{6}) \hat{S}|0}.  \nonumber
\end{eqnarray}
In order to evaluate Eq.~(\ref{eq:thpnc}) we determine the expectation value
\begin{eqnarray}
\label{eq:hpnceval}
&&\hspace{-5ex}\braket{\hat{a}^{\dagger}_{i}(\omega_{1})\hat{a}^{\dagger}_{s}(\omega_{2}) \hat{a}^{\dagger}_{s}(\omega_{3}) \hat{a}_{s}(\omega_{4}) \hat{a}_{s}(\omega_{5})  \hat{a}_{i}(\omega_{6})}  
= \big< \hat{a}_{s}(\omega_{p}-\omega_{1})  \alpha^{*}(\omega_{p}-\omega_{1}) \big [ \hat{a}^{\dagger}_{s}(\omega_{2})\beta(\omega_{2}) - \hat{a}_{i}(\omega_{p}-\omega_{2})\alpha^{*}(\omega_{2})\big] \nonumber \\
&&\hspace{33ex}\otimes \big[ \hat{a}^{\dagger}_{s}(\omega_{3})\beta(\omega_{3}) - \hat{a}_{i}(\omega_{p}-\omega_{3})\alpha^{*}(\omega_{3})\big]
\big[ \hat{a}_{s}(\omega_{4})\beta(\omega_{4}) - \hat{a}^{\dagger}_{i}(\omega_{p}-\omega_{4})\alpha(\omega_{4})\big] \nonumber \\
&&\hspace{33ex} \otimes\big[ \hat{a}_{s}(\omega_{5})\beta(\omega_{5}) - \hat{a}^{\dagger}_{i}(\omega_{p}-\omega_{5})\alpha(\omega_{5})\big] \hat{a}^{\dagger}_{s}(\omega_{p}-\omega_{6})  \alpha(\omega_{p}-\omega_{6}) \big> \nonumber \\
&=&\hspace{-2ex} \alpha^{*}(\omega_{p}-\omega_{1})\beta(\omega_{2}) \alpha^{*}(\omega_{3})\beta(\omega_{4})\alpha(\omega_{5})   \alpha(\omega_{p}-\omega_{6}) 
\underbrace{\braket{\hat{a}_{s}(\omega_{p}-\omega_{1}) \hat{a}^{\dagger}_{s}(\omega_{2})\hat{a}_{i}(\omega_{p}-\omega_{3}) \hat{a}_{s}(\omega_{4})\hat{a}^{\dagger}_{i}(\omega_{p}-\omega_{5}) \hat{a}^{\dagger}_{s}(\omega_{p}-\omega_{6}) }}_{=\delta(\omega_{p}-\omega_{6}-\omega_{4})\delta(\omega_{3}-\omega_{5})\delta(\omega_{p}-\omega_{1}-\omega_{2})} \nonumber  \\
&+& \hspace{-2ex} \alpha^{*}(\omega_{p}-\omega_{1})\beta(\omega_{2}) \alpha^{*}(\omega_{3})\alpha(\omega_{4})\beta(\omega_{5})   \alpha(\omega_{p}-\omega_{6}) 
\underbrace{\braket{\hat{a}_{s}(\omega_{p}-\omega_{1}) \hat{a}^{\dagger}_{s}(\omega_{2})\hat{a}_{i}(\omega_{p}-\omega_{3}) \hat{a}^{\dagger}_{i}(\omega_{p}-\omega_{4})\hat{a}_{s}(\omega_{5}) \hat{a}^{\dagger}_{s}(\omega_{p}-\omega_{6}) }}_{=\delta(\omega_{p}-\omega_{6}-\omega_{5})\delta(\omega_{3}-\omega_{4})\delta(\omega_{p}-\omega_{1}-\omega_{2})} \nonumber \\
&+&\hspace{-2ex}  \alpha^{*}(\omega_{p}-\omega_{1})\alpha^{*}(\omega_{2}) \beta(\omega_{3})\beta(\omega_{4})\alpha(\omega_{5})   \alpha(\omega_{p}-\omega_{6}) \underbrace{\braket{\hat{a}_{s}(\omega_{p}-\omega_{1}) \hat{a}_{i}(\omega_{p}-\omega_{2})\hat{a}^{\dagger}_{s}(\omega_{3}) \hat{a}_{s}(\omega_{4})\hat{a}^{\dagger}_{i}(\omega_{p}-\omega_{5}) \hat{a}^{\dagger}_{s}(\omega_{p}-\omega_{6}) }}_{=\delta(\omega_{p}-\omega_{6}-\omega_{4})\delta(\omega_{2}-\omega_{5})\delta(\omega_{p}-\omega_{1}-\omega_{3})} \nonumber \\
&+&\hspace{-2ex}  \alpha^{*}(\omega_{p}-\omega_{1})\alpha^{*}(\omega_{2}) \beta(\omega_{3})\alpha(\omega_{4})\beta(\omega_{5})   \alpha(\omega_{p}-\omega_{6}) \underbrace{\braket{\hat{a}_{s}(\omega_{p}-\omega_{1}) \hat{a}_{i}(\omega_{p}-\omega_{2})\hat{a}^{\dagger}_{s}(\omega_{3}) \hat{a}^{\dagger}_{i}(\omega_{p}-\omega_{4})\hat{a}_{s}(\omega_{5}) \hat{a}^{\dagger}_{s}(\omega_{p}-\omega_{6}) }}_{=\delta(\omega_{p}-\omega_{6}-\omega_{5})\delta(\omega_{2}-\omega_{4})\delta(\omega_{p}-\omega_{1}-\omega_{3})} \nonumber \\
&+&\hspace{-2ex}  \alpha^{*}(\omega_{p}-\omega_{1})\alpha^{*}(\omega_{2}) \alpha^{*}(\omega_{3})\alpha(\omega_{4})\alpha(\omega_{5})   \alpha(\omega_{p}-\omega_{6}) \nonumber\\
&&\hspace{24ex}\cdot\underbrace{\braket{\hat{a}_{s}(\omega_{p}-\omega_{1}) \hat{a}_{i}(\omega_{p}-\omega_{2})\hat{a}_{i}(\omega_{p}-\omega_{3}) \hat{a}^{\dagger}_{i}(\omega_{p}-\omega_{4})\hat{a}^{\dagger}_{i}(\omega_{p}-\omega_{5}) \hat{a}^{\dagger}_{s}(\omega_{p}-\omega_{6}) }}_{=\delta(\omega_{6}-\omega_{1})[\delta(\omega_{3}-\omega_{4})\delta(\omega_{2}-\omega_{5})+\delta(\omega_{2}-\omega_{4})\delta(\omega_{3}-\omega_{5})]},
\end{eqnarray}
in the final form of which only few terms survive. Plugging Eq.~(\ref{eq:hpnceval}) in Eq.~(\ref{eq:thpnc}) results in Eq.~(\ref{eq:III}).

\end{document}